\documentclass{nature}
\usepackage[T1]{fontenc}
\usepackage{float}
\usepackage{amsmath}
\usepackage{amssymb}
\usepackage {mathtools}
\usepackage{indentfirst}
\usepackage{textcomp}
\usepackage{color}
\usepackage{soul}
\usepackage{ulem}
\usepackage{siunitx}
\usepackage{amsmath}
\usepackage{graphicx}
\usepackage{caption}
\usepackage[version=4]{mhchem}

\usepackage{multirow}
\usepackage{array}
\usepackage{booktabs} 
\usepackage{lineno}
\usepackage[inline]{showlabels}

\usepackage{xr}
\makeatletter
\newcommand*{\addFileDependency}[1]{
  \typeout{(#1)}
  \@addtofilelist{#1}
  \IfFileExists{#1}{}{\typeout{No file #1.}}
}
\makeatother

\title{Quantum-activated neural reservoirs on-chip open up large hardware security models for resilient authentication}

\author{Zhao He$^{1}$\textsuperscript{\textdagger}, Maxim S. Elizarov$^1$\textsuperscript{\textdagger}, Ning Li$^1$, Fei Xiang$^1$, and Andrea Fratalocchi$^1$}

\begin{document}
\maketitle

\begin{affiliations}
 \item PRIMALIGHT, Faculty of Electrical Engineering, King Abdullah University of Science and Technology (KAUST), Thuwal 23955-6900, Saudi Arabia. \\ \textsuperscript{\textdagger}first authors with equal contribution
\end{affiliations}
\begin{abstract}
Quantum artificial intelligence is a frontier of artificial intelligence research, pioneering quantum AI-powered circuits to address problems beyond the reach of deep learning with classical architectures. This work implements a large-scale quantum-activated recurrent neural network possessing more than $3$~trillion hardware nodes/cm$^2$, originating from repeatable atomic-scale nucleation dynamics in an amorphous material integrated on-chip, controlled with 0.07~nW electric power per readout channel. Compared to the best-performing reservoirs currently reported, this implementation increases the scale of the network by two orders of magnitude and reduces the power consumption by six, reaching power efficiencies in the range of the human brain, dissipating 0.2~nW/neuron. When interrogated by a classical input, the chip implements a large-scale hardware security model, enabling dictionary-free authentication secure against statistical inference attacks, including AI's present and future development, even for an adversary with a copy of all the classical components available. Experimental tests report 99.6\% reliability, 100\% user authentication accuracy, and an ideal 50\% key uniqueness. Due to its quantum nature, the chip supports a bit density per feature size area three times higher than the best technology available, with the capacity to store more than $\mathbf{2^{1104}}$ keys in a footprint of $1$~ cm$^2$. Such a quantum-powered platform could help counteract the emerging form of warfare led by the cybercrime industry in breaching authentication to target small to large-scale facilities, from private users to intelligent energy grids.
\end{abstract}

\section*{Introduction}
\noindent 
Reservoir computers are recurrent neural networks comprising a highly-dimensional untrained network encoding dynamical patterns of feature spaces that a trained readout decodes into desired outputs~\cite{cucchi2022hands}. Reservoirs implemented using classical systems reported highly performing applications in dynamic vision \cite{tan2023dynamic}, temporal data classification \cite{moon2019temporal}, pattern recognition \cite{abreu2020role,korber2023pattern}, stock market forecasting \cite{liu2022forecasting}, and language processing~\cite{valensise2022large}. Research on this technology attracts interest for the potential to offer energy-efficient hardware training to large-scale neural models, including modern transformer architectures and foundation models~\cite{bommasani2021opportunities}. The current state of the art comprises large-scale models for optical processors~\cite{valensise2022large} and brain organoids~\cite{cai2023brain} with a typical neural size of $10^7$ to $10^{10}$ hardware nodes, dissipating powers in the range between mW to $\mu$W per readout channel. Pioneering theoretical work suggests that quantum mechanics could offer a way to upgrade performance, opening neural models that could enhance applications beyond the reach of classical architectures~\cite{ghosh2021realising}.\\ 
Here, we implement a large-scale quantum neural reservoir (QNR) on a chip, leveraging a physical network activated by quantum nucleations on a suitably engineered amorphous phase change material film. The characteristic size of each node element, probed with an atomic force microscope (AFM), has a spatial area of $8$~nm~$\times$~$8$~nm and allows a QNR with 1~cm$^2$ footprint to support more than $3\times 10^{12}$ nodes.
The large-scale nature of this network, combined with its quantum nature, opens significant application opportunities in information security. This area is recognized of global importance as modern society moves towards cloud computing, 6G, smart energy grids, and the Internet of Things (IoT), with the transmission of more than three quintillion bytes every day~\cite{naeem2022trends,adam2001cryptography,li2017robust,joshi2020trusted}. Traditional methods such as passwords and biometrics are no longer adequate to protect users and have led to documented high-profile security breaches~\cite{robbins2018once,misra2019lesson,neto2021developing}, including the largest biometric identity database in the world~\cite{misra2019lesson}, with hundreds of millions of users affected~\cite{robbins2018once,neto2021developing}.\\
Research is currently investing significant efforts to address this issue with more robust hardware security primitives, stimulating work in different materials and configurations, ranging from carbon nanotubes~\cite{zhong2022twin}, to biological species~\cite{wali2019biological}, smart inks \cite{moglianetti2022nanocatalyst}, silicon photonics~\cite{wali2019biological,hu2016physically,dodda2021graphene,john2021halide,DiFalco2019,fratalocchi2021nist,liu2019inkjet,camilleri2023using,roberts2015using,fukuoka2022physically}, and integrated electronics~\cite{clarke2010authentication,camilleri2023using,roberts2015using}. On the hardware side, the main challenge is developing a platform that is not affected by nanotechnology development, which demonstrated successful cloning of complex physical primitives previously considered unclonable~\cite{helfmeier2013cloning,oren2013effectiveness}. On the software side, the issue is addressing artificial intelligence \cite{talukder2021memory}. Because a deterministic input-output function represents the security primitives' response, statistical learning can infer the function from the information flow over the network and clone the system. This problem affects all current technologies that transmit data through classical channels~\cite{ruhrmair2010modeling,saha2013model}. Although researchers have reported primitives that resist selected known AI attacks~\cite{jin2023low,sajadi2023dc,wang2020lattice,nguyen2017mxpuf}, there is not yet a technology demonstrated secure against any possible inference technique, including future or not publicly disclosed development. We here demonstrate that the QNR chip can address these issues with a technology protected against hardware cloning by the laws of quantum mechanics and immune to statistical inference, offering a platform that could help support future milestones in the digitalization of modern society toward smart cities with deep integration of critical infrastructure technologies.

\section*{Results}
\subsection{Implementation of a quantum neural reservoir (QNR) on-chip}
\begin{figure*}
\centering
\includegraphics[width=\textwidth]{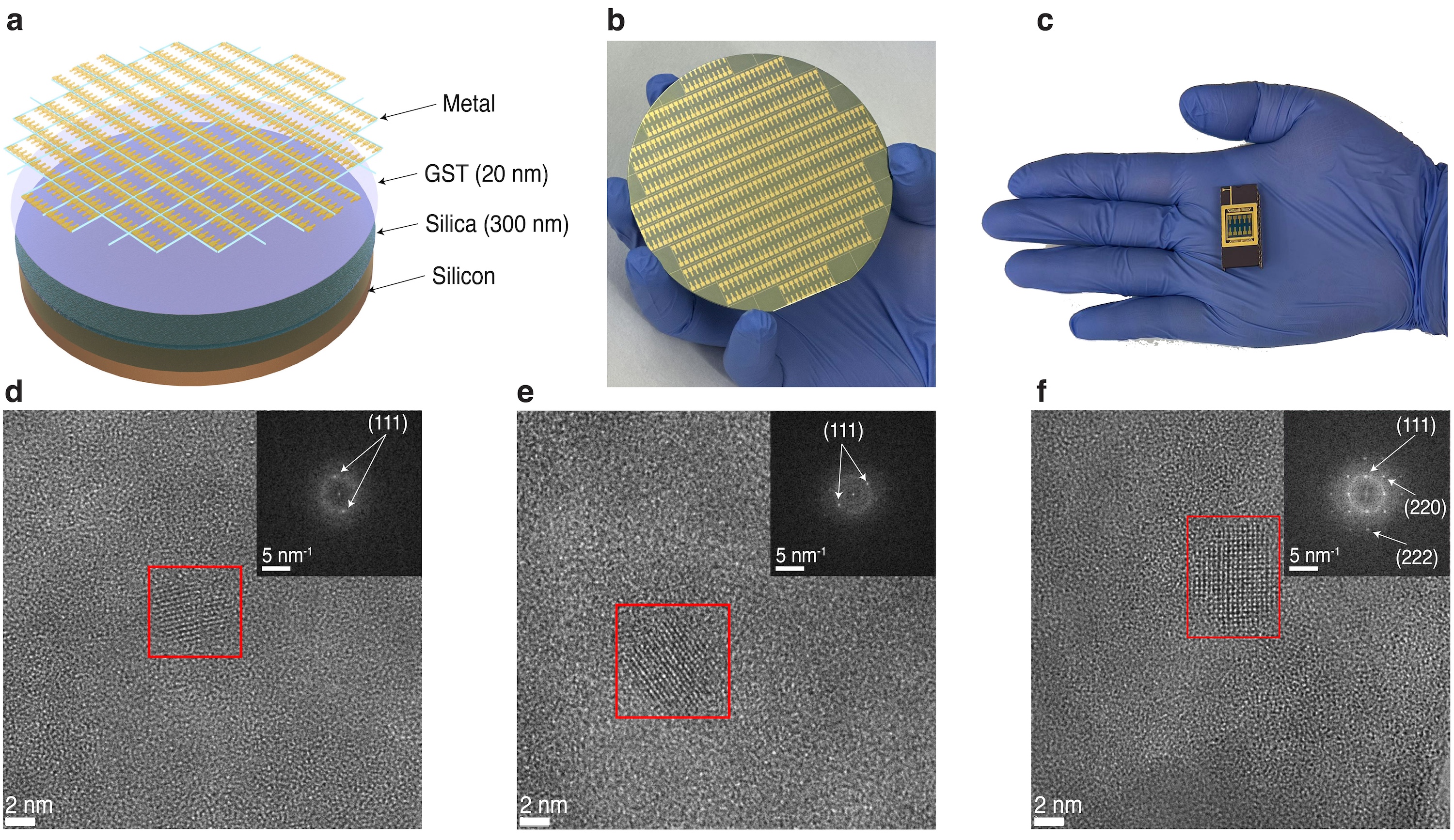}
\caption{\textbf{QNR implementation and TEM characterization} (a) Schematic diagram of a QNR chipset covering a 4-inch wafer area; (b) Image of manufactured chips on the wafer; (c) Picture of single QNR device wire bounded to a side-brazed dual in-line ceramic package; (d)-(f) HRTEM images of different regions of the as-deposited GST film on the chip and corresponding SAED patterns of regions with nanocrystalline phase.}
\label{Fabrication_process}
\end{figure*}

\begin{figure*}
\centering
\includegraphics[width=0.99\textwidth]{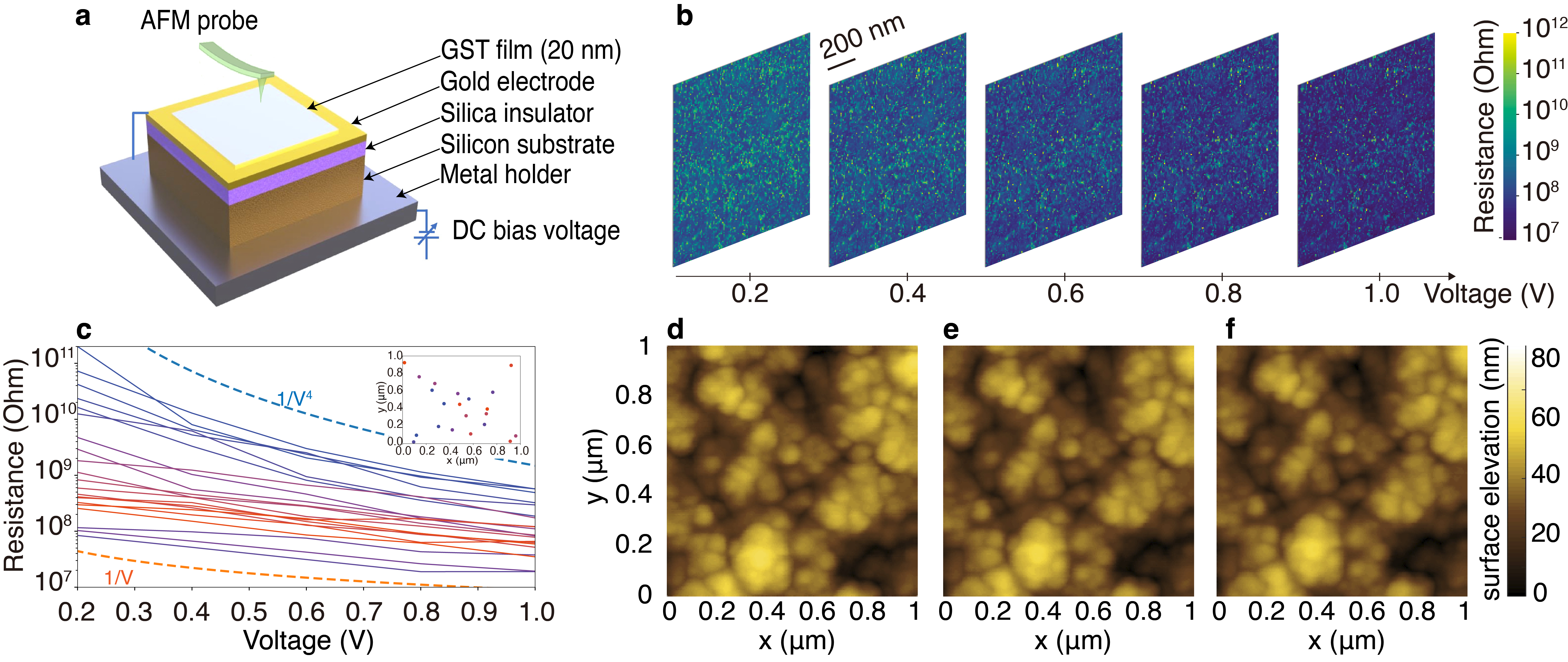}
\caption{\textbf{AFM measurement results} (a) Schematic diagram of the device's setup; (b) Point-to-point Resistance-Voltage spatial maps of a (1\(\ \mu \)m$\times$1\(\ \mu \)m) chip area obtained at increasing applied bias $V$; (c) Resistance-Voltage curves (solid lines) of selected set points in the chosen area (inset, colored markers) for applied bias $V$ between 0.2V and 1V. The color of each line indicates the evolution of the point marked with the same color in the inset; (d-f) Topography images of the selected area of the different applied bias $V$.}
\label{AFM}

\end{figure*}

\begin{figure*}
\centering
\includegraphics[width=0.99\textwidth]{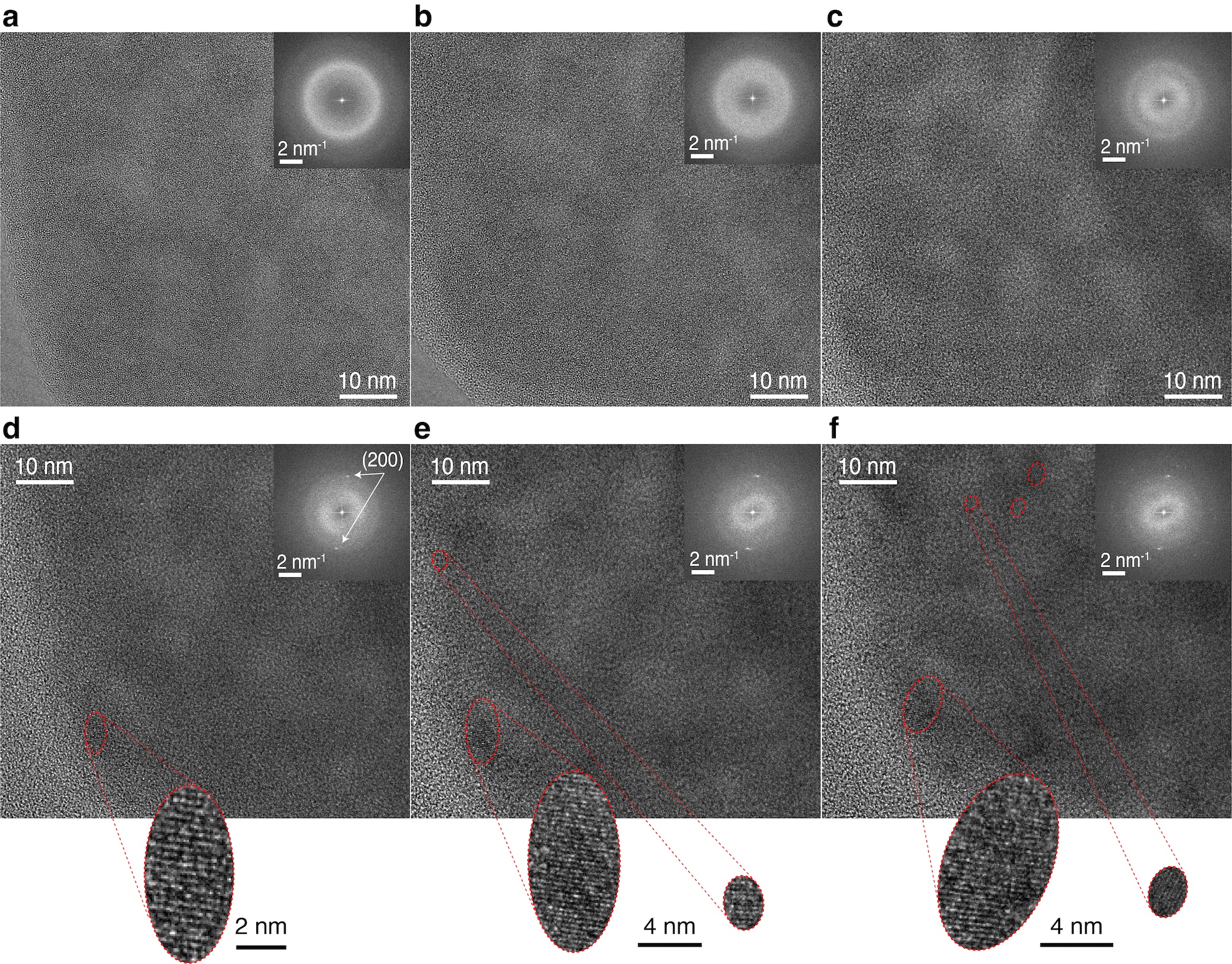}
\caption{\textbf{High-Resolution TEM results} HRTEM images and corresponding SAED patterns of GST film treated at (a) Room temperature, (b) 75 $^\circ C$, (c) 125 $^\circ C$, (d) 150 $^\circ C$, (e) 160 $^\circ C$, and (f) 170 $^\circ C$. We highlighted in red the portion of the sample showing quantum nucleated areas with crystalline phases.}
\label{TEM_image}

\end{figure*}

\begin{figure*}
\centering
\includegraphics[width=0.99\textwidth]{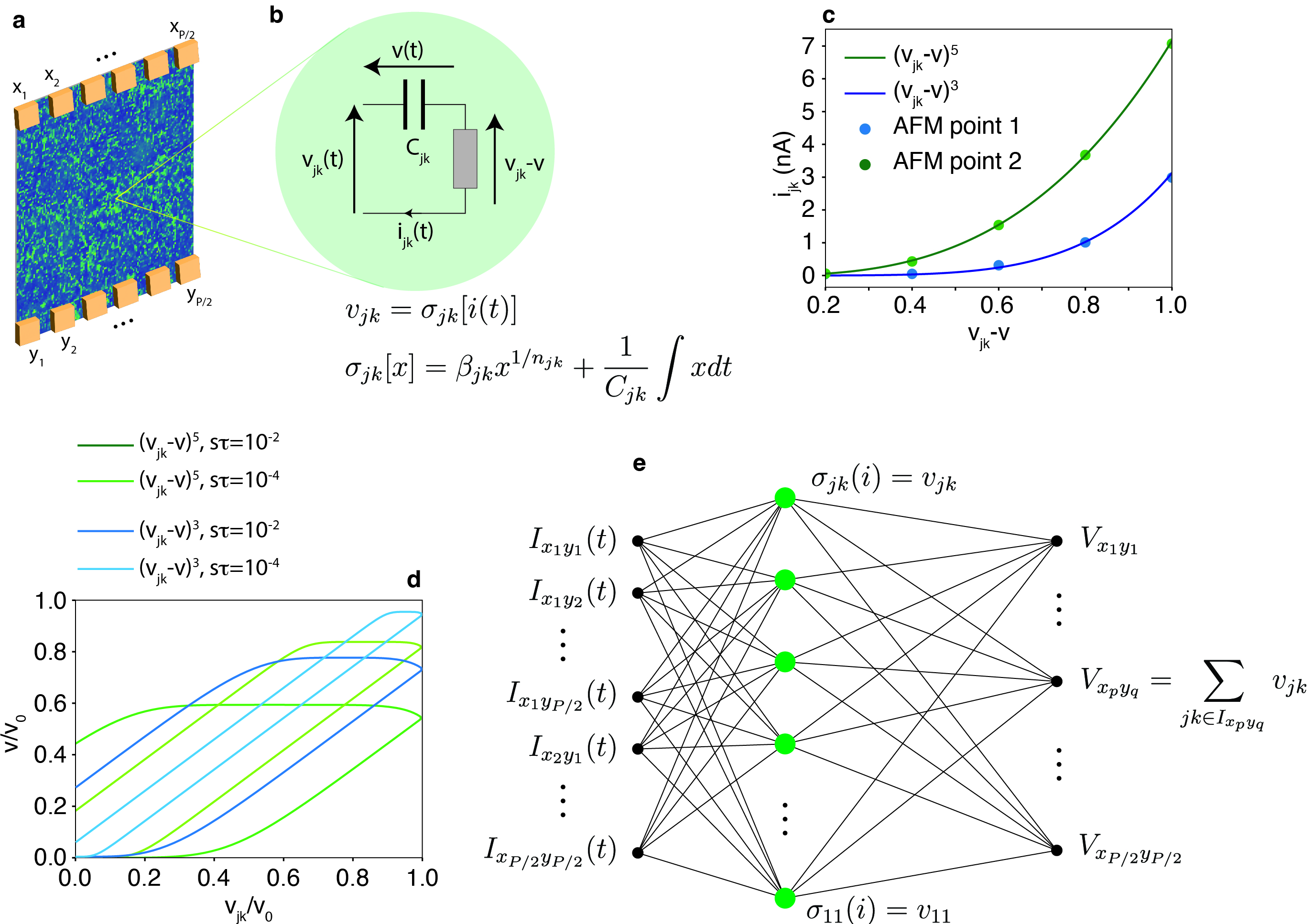}
\caption{\textbf{Data-drive model of the QNR response} (a) Schematic diagram of the QNR chip with a set of input $x_p$ and output $y_q$ electrodes applied to the amorphous phase change material with random nonlinear resistivity distribution versus applied voltage (colormap) measured in Fig.\ref{AFM}. (b) Nanoscale circuit equivalent to a single quantum-activated nanocircuit of the QNR of $8$~nm~$\times$~$8$~nm area; (c) Current response of two randomly selected points in the AFM map and their equivalent nonlinear current-voltage distribution retrieved from AFM ( markers) and theoretical curve (solid line). (d) Simulation results on a single nanocircuit of the panel (b) for a triangular input excitation and different sweep rates. (e) Network structure of the entire system of quantum-activated nanocircuits.}
\label{theory}

\end{figure*}

Figure \ref{Fabrication_process}a shows a schematic diagram of the QNR module. The system comprises a nanolayer of \ce{Ge2Sb2Te5} (GST) material on a silica insulator platform grown on a silicon wafer, completed by metallic contacts on the top side. GST is a phase change material possessing nanosecond switching between amorphous and crystalline phases triggered by temperature increase applied directly or indirectly through a current flow~\cite{loke2012breaking}. When the phase changes, the material exhibits several orders of magnitude difference in resistivity~\cite{ gholipour2019promise}, and a factor of two in refractive index~\cite{hosseini2014optoelectronic}.\\
Module manufacturing uses a scalable CMOS-compatible process on 4-inch wafers (Fig.~\ref{Fabrication_process}a-c). The process grows a 300~nm thick silica insulator film generated on a silicon wafer by thermal oxidation, followed by a 20~nm thick GST film suitably deposited by RF magnetron sputtering, and completed by 200 nm thick electrode patterns obtained by DC magnetron sputtering and photo-lithography (Fig.~\ref{Fabrication_process}a). Figure~\ref{Fabrication_process}b-c shows the photo of a final integrated chip with ten electrodes wire bonded to a side-brazed dual in-line ceramic package (see Methods). 
Figure~\ref{Fabrication_process}d-f shows high-resolution transmission electron microscopy (HRTEM) images of three different sections of the as-deposited GST film with far-field diffraction patterns. The photos show a microscopic phase of a large amorphous region with several crystalline areas of 5-7~nm characteristic dimension. The far-field recorded patterns in Fig.~\ref{Fabrication_process}d-f show that the crystalline quantum grains are randomly located and oriented in space. The Brillouin patterns visualize various crystallographic planes of the face-centered cubic (FCC) lattice of the GST crystalline phase. Different material areas present multiple crystalline planes and diffraction rings emerging from the far-field spectrum, indicating the presence of polycrystalline structures.\\
Figure~\ref{AFM} characterizes the electrical properties at the nanoscale of the device by Atomic Force Microscopy (AFM). Figure~\ref{AFM}a shows the schematic diagram of the sample structure probed by AFM. The system comprises the growth of a \SI{2}{\micro\metre} silica insulator film on a silicon substrate, followed by the deposition of 195~nm thick gold bottom electrode film and deposition of 20~nm thick GST film on the top surface of the sample (see Methods).\\
Figure~\ref{AFM}b displays the results of the GST spatial distribution of resistance represented in a selected area of 1\(\ \mu \)m$\times$1\(\ \mu \)m, obtained with an ultra-sharp AFM probe tip with a radius of $25$~nm and spatial sensitivity of $8$~nm along the lateral plane. The resistance of the sample shows large fluctuations from point to point of various orders of magnitude, providing speckle-like distributions at different applied biases. The coherent area of each speckle, calculated from the autocorrelation of the spatial maps, is a single pixel with an area of 8~nm$\times$8~nm.\\ 
Figure~\ref{AFM}c plots the single point nanoscale resistance $R$ versus applied bias $V$ measured at a random distribution of spatial points (Fig.~\ref{AFM}c, inset). When the voltage $V$ increases, the resistance $R$ decreases by one to four orders of magnitude, following power law behaviors $R(V)=\sum \alpha_n V^{-n}$ with microscopic evolution changing from point to point. Resistance values, which vary between $10^{12}~\Omega$ and $10^7~\Omega$, indicate that the material shows a progressive tendency to change phase from amorphous, where the GST is insulative, to crystalline, where the material is conductive~\cite{nevzorov2023two}. Figure~\ref{AFM}d-f provides AFM topography images obtained with increasing applied voltage to (d) 0~V, (e) 0.6~V, and (f) 1~V. The photos show no appreciable differences. This result implies that the random and nonlinear resistance of the material (Fig.~\ref{AFM}b-c) originates from nanoscale dynamics occurring at lower spatial scales than AFM's resolution.\\
Figure~\ref{TEM_image} illustrates the in situ HRTEM imaging results to study this process. In these measurements, we employ a smart chip on the HRTEM sample grid to set a constant temperature for heating the sample, allowing observation of the nanoscale details of the material phase transition (see Methods). We vary the temperature from room temperature (Figure~\ref{TEM_image}a) to 170 degrees Celsius (Fig.~\ref{TEM_image}f) with step intervals at 75$^\circ C$ (Fig.~\ref{TEM_image}b), 125$^\circ C$ (Fig.~\ref{TEM_image}c), 150$^\circ C$ (Fig.~\ref{TEM_image}d), and 160$^\circ C$ (Fig.~\ref{TEM_image}e). We image the spatial material distribution and selected area electron diffraction (SAED) patterns at each temperature. The results of Fig.~\ref{TEM_image} show that when the temperature increases, two different mechanisms of material change occur in the GST film. First, the amorphous region of the material changes its properties. The HRTEM images show this evolution quantitatively, with the appearance of larger aggregates of randomly oriented grains (Fig.~\ref{TEM_image}a towards f). In this dynamical evolution, the SAED patterns appear as far-field diffraction rings with varying intensities and shapes, corresponding to scattered responses from different amorphous material structures.
In parallel with this process, the GST film nucleates quantum-sized crystals (Fig.~\ref{TEM_image}d-f). The indexed SAED patterns in Fig.~\ref{TEM_image}d-f confirm this point by showing the characteristics of face-centered cubic (FCC) structures of GST in the crystalline phase. Nanocrystals nucleate at random positions, with characteristic areas gradually increasing with temperature from a few atoms size (Fig.~\ref{TEM_image}e-f magnified areas), with the progressive appearance of sharper diffraction peaks in SAED patterns (Fig.~\ref{TEM_image}d-f).\\
Figure~\ref{theory} combines the AFM and in situ HRTEM results of Figs.~\ref{AFM}-\ref{TEM_image}, providing a data-driven theoretical model of the system. When current flows between a combination of electrodes $x_iy_j$ in the device (Fig.~\ref{theory}a), charge carriers travel inside the amorphous GST phase, inducing quantum grains nucleation sustained from the induced bias variation and observed as nonlinear resistivity behaviors (Fig.~\ref{AFM}). At a single spatial point, the material response is an equivalent nanoscale circuit comprising two elements (Fig.\ref{theory}b): a nonlinear element (Fig.~\ref{theory} gray area), which models the nonlinear resistivity measured in Fig.~\ref{AFM}b-c, in series with a nanoscale capacitor representing the material capacitance. Because GST capacitance does not exhibit a significant variation between the amorphous and crystalline phases~\cite{ghamlouche2007capacitance}, the model considers the leading contribution of voltage-independent capacitance. Figure~\ref{theory}c shows the current response of two randomly selected points in the AFM map and their equivalent nonlinear current-voltage distribution fitted from experimental data (circle markers). Because the quantum nucleation of the GST occurs randomly in space (Fig.~\ref{TEM_image} and Fig.~\ref{theory}d on one AFM pixel area), each spatial point experiences a different microscopic phase change process and exhibits a different nonlinear response with diverse power law behavior.\\
Figure~\ref{theory}d shows simulation results on a single nanocircuit of Fig.~\ref{theory}b using a triangular input excitation with different sweep rates (see Methods). The nanoscale circuit provides hysteretic behavior, generating smaller loops with lower acquired capacitance as the sweep rates decrease. At lower sweep rates, the time variation of $|\dot v|$ in the capacitor follows the input $|\dot v_{jk}|$ and $v_{jk}\approx v$, thus inhibiting the nonlinearity of the material with $v_{jk}-v=v_r\approx 0$. Faster sweep rates, in contrast, generate more significant differences between $|\dot v|$ and $|\dot v_{jk}|$, driving a stronger material nonlinearity in $v_r$ and creating larger hysteresis-like loops.\\
When electron carriers flow through the chip from a combination of input-output $x_p-y_q$ electrodes, electrons drift in space, generating nanoscale voltage variations at each spatial point $v_{jk}(t)$ following the effects of the quantum nucleation-driven circuit of Fig.~\ref{theory}b-d. Indicating with $i_{jk}(t)$ the current flow at spatial point $i,j$, the voltage generated $v_{jk}(t)$ is:
\begin{equation}
    \label{eq0}
    v_{jk}=\sigma_{jk}(i)=\beta_{jk}i_{jk}^{1/n_{jk}}+\frac{1}{C_{jk}}\int i_{jk}(t) dt,
\end{equation}
with $\beta_{jk}$ and $n_{jk}$ parameters modeling the point-to-point nonlinear material response (Fig.~\ref{AFM} and Fig.~\ref{theory}c), and $C_{jk}$ the nanoscale capacitance. The dynamics of Eq.~\eqref{eq0} is equivalent to the activation function of a recurrent neural network, which processes the current arising from each electrode pair excited at the input with a nonlinear transmission function $v_{jk}=\sigma(i)$ (Fig.~\ref{theory}e). The activation function of each spatial node contains a non-polynomial instantaneous $\propto i_{jk}^{1/n_{jk}}$ term, satisfying the condition of the universal approximation theorem of neural networks~\cite{Hornik1989MultilayerFN}. The nonlocal contribution $\int i_{jk}(t) dt$ term provides a fading memory effect~\cite{cucchi2022hands} that processes past network outputs. For the specific input selected in Fig.~\ref{theory}d, the fading memory at each neural node manifests itself as hysteretic loops. The voltage variations $v_{jk}(t)$ acquired by the electrons in their spatial travel path $i,j\in\mathcal{I}_{x_py_q}$ sum up in an equivalent readout layer that outputs a voltage signal from the chip (Fig.~\ref{theory}d). The nodes are accessible for training in the readout via input-output electrodes placed on the surface of the chip.

\subsection{Application in resilient authentication}
\begin{figure*}
\centering
\includegraphics[width=0.99\textwidth]{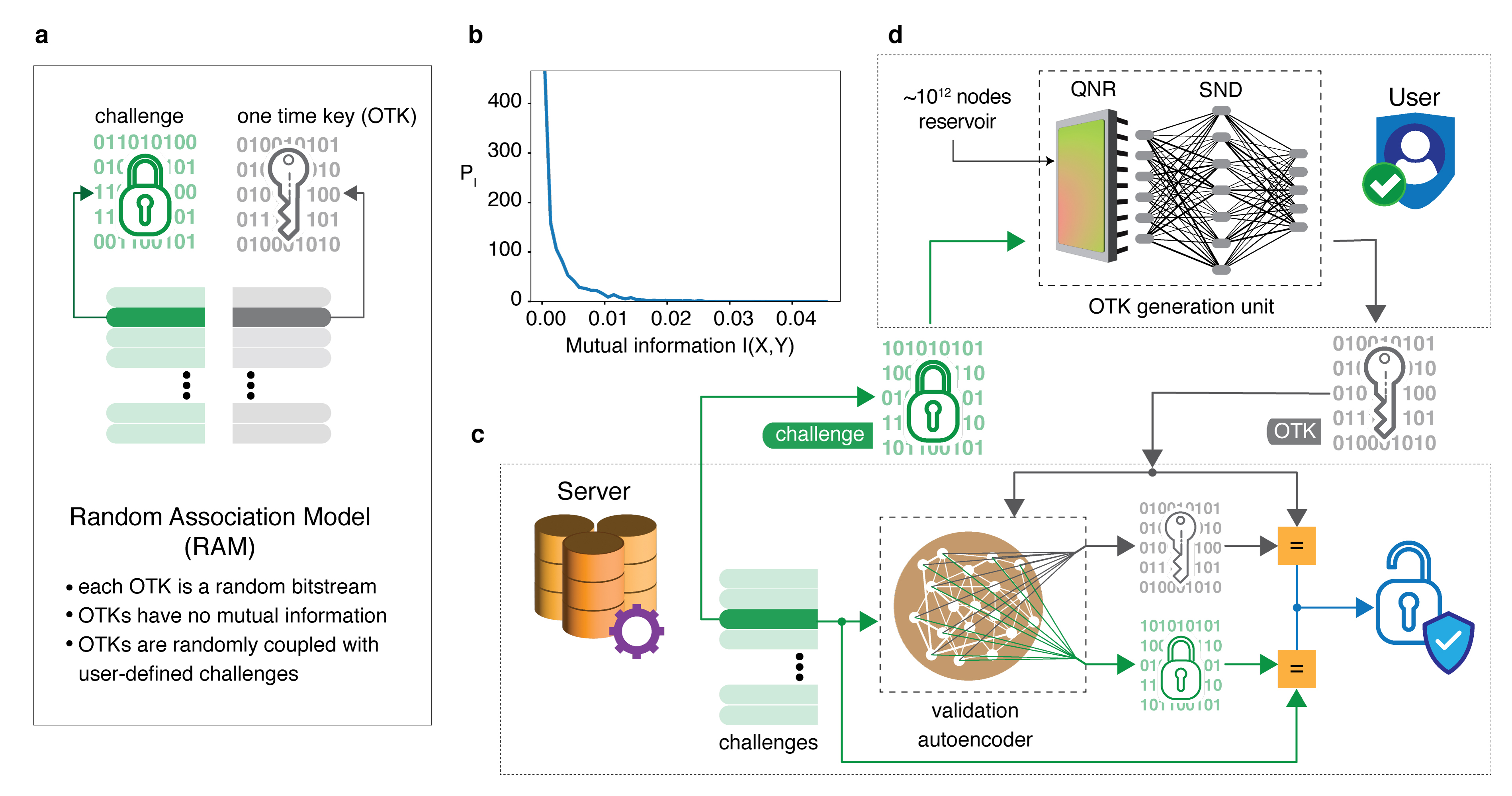}
\caption{\textbf{proposed architectures of QNR-based user authentication pipeline} (a) A RAM creates random associations between one-time keys (OTK) with no mutual information with each other and a set of challenges. (b) Probability distribution of mutual information in a set of 10000 keys created from physical noise. (c) Server authentication model and challenge-response authentication steps. The server challenges the hardware security primitive (d), comprising the QNR with a nonlinear readout decoder unit, authenticating the received key with a validation autoencoder (VA). The VA does not use a dictionary or store any challenge-response relationship in the RAM.}
\label{concept_image1}

\end{figure*}

Figure~\ref{concept_image1} summarizes the high-level structure of a resilient authentication scheme that utilizes the large-scale network size of the QNR chip. The system uses a random association model (RAM) to guarantee the impossibility of any form of inference from the data flow over the network (Fig.~\ref{concept_image1}a). The RAM provides a stochastic random association between arbitrarily chosen user-defined challenges and one-time keys (OTK). Each OTK is a random bit sequence created under the condition that any single key has no mutual information (MI) entropy~\cite{10.5555/1146355} with the set of previously generated keys. The mutual information (MI) between a bit $X=(0,1)$ of a key and a bit $Y=(y_0,y_1)=(0,1)$ of any other is defined as $I(X,Y)=\sum_{i,j}p(x_i,y_j)\log\frac{p(x_i,x_j)}{p(x_i)p(y_j)}$, with $p(x_i,y_j)$ joint and $p(x_i)$, $p(y_i)$ marginal probabilities of bits $X$ and $Y$ belonging to two different OTK. This quantity measures the information gained from the knowledge of the bits of one OTK bit sequence when predicting another. The disappearance of mutual information implies that knowing one sequence does not reduce the uncertainty of another~\cite{10.5555/1146355}. Under this condition, an adversary who intercepts the data flowing on the network, including all challenges and OTKs, and attempts to infer the new key cannot do better than best guess it without seeing the prior OTKs. The reason is that a given set of $n-1$ keys can generate the next $n$ from a pool of infinite combinations without relation to the initial set. The arbitrariness of this choice makes it impossible to predict the chosen one with certainty.\\
Figure~\ref{concept_image1}b illustrates an implementation of this concept by computing the MI in a set of 10000 OTKs, obtained by selecting 256-bit long random bitstreams from five independent sequences of physical noise~\cite{fratalocchi2021nist}. OTK bitstreams have information gains below $5\times 10^{-3}$. Developing a software model that infers the next OTKs in this process is impossible because the future choice is independent of the past, which provides no information gain to reduce uncertainty in future decisions.\\
In the scheme of Fig.~\ref{concept_image1}, no party stores the RAM. The server retains only the set of user-defined challenges (Fig.~\ref{concept_image1}c). It generates OTKs on the user side (Fig.~\ref{concept_image1}d) by using the QNR chip and a software neural decoder (SND) readout implemented with a feedforward neural network. The reservoir, randomly initialized at the atomic level during manufacturing, autonomously evolves its unique nuclear structure. When interrogated by an input challenge, the QNR encodes the challenge into a complex dynamical orbit, which a trained SND decodes into the corresponding OTK.\\
One-time authentication starts from the server following a challenge-response model~\cite{10.5555/2412064}: the server randomly chooses a challenge, sends it to the user, and receives the response as an OTK. Unlike traditional architectures, the server does not use a dictionary database; it employs a validation autoencoder (VA) to verify the OTK and authenticate the user (Fig.~\ref{concept_image1}c). The VA comprises a neural network unit trained to output the OTK and the associated challenge only if the OTK-challenge pair sent at the input belongs to the RAM. Conversely, if the input differs from an OTK-challenge pair, the autoencoder outputs a noisy bit sequence without mutual information. The system authenticates the user if the autoencoder output is identical to the OTK sent at the input and the challenge that initiates the authentication (Fig.~\ref{concept_image1}c). Because the sole functionality learned in the autoencoder is to replicate a discrete number of binary input sequences at the output, the system of Fig.~\ref{concept_image1}c never memorizes any challenge-response relation in the RAM.

\subsection{Key generation} 
\begin{figure*}
\centering
\includegraphics[width=0.99\textwidth]{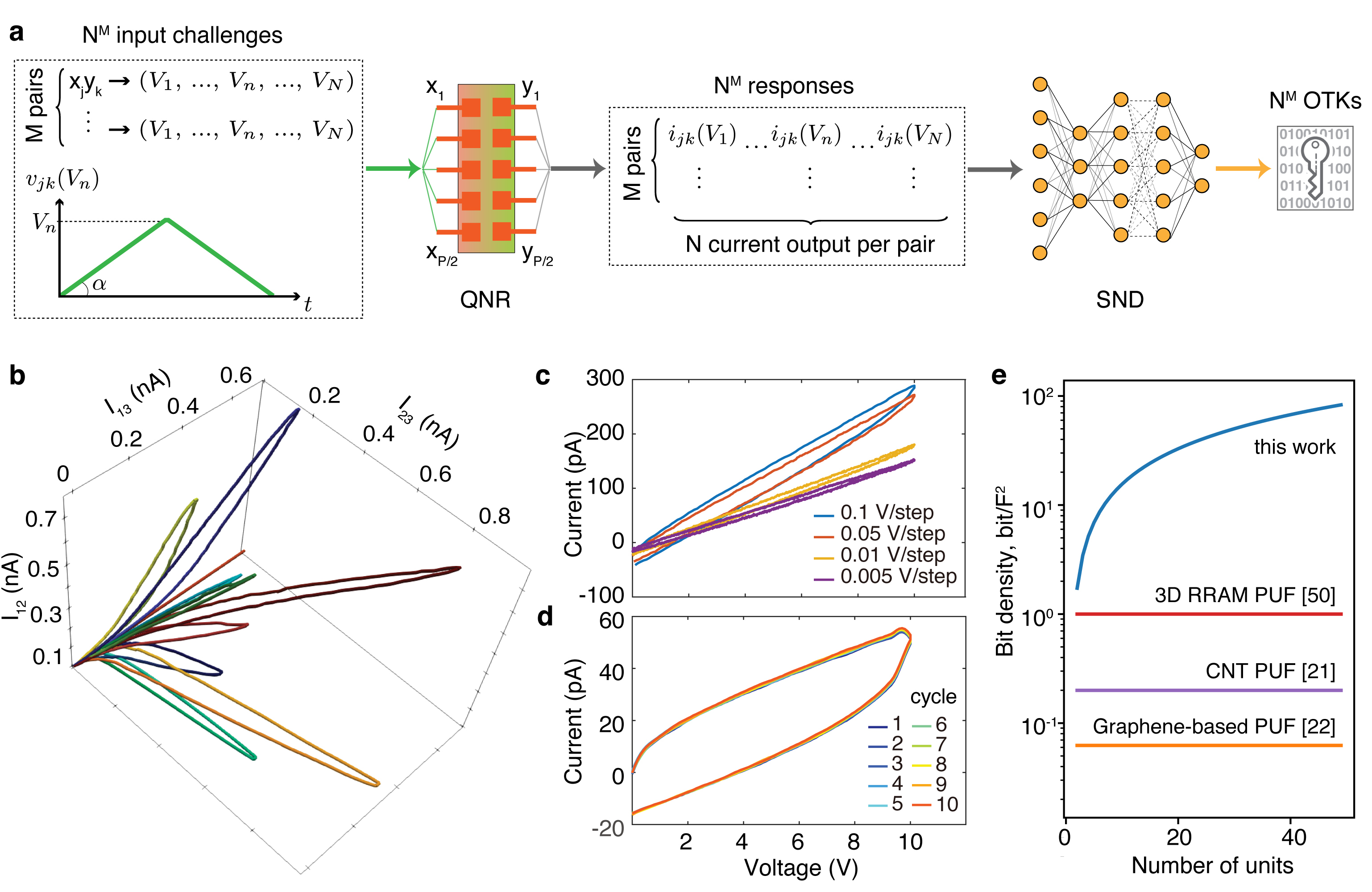}
\caption{ \textbf{QNR key generation and performance analysis} (a) OTK generation by QNR encoder and software neural decoder (SND) for a given input challenge. The process comprises a set of voltages $V_n$ to construct triangular waveforms applied to the input electrodes. This results in complex dynamical current orbits' evolution in time, which the SND decodes into OTKs. (b) Experimental multidimensional dynamical orbits retrieved for a space of 10 selected challenges; (c-d) Corresponding Current-Voltage curves for (c) different voltage sweep rates and (d) the same sweep rate and ten consecutive measurements; (e) State-of-the-art figure of merit comparing existing technology using the manufacturing independent metric of bit density per feature size area $bit/F^2$.}
\label{keygen}

\end{figure*}

Figure \ref{keygen} summarizes experimental and theoretical results on OTK generation with the QNR. Figure \ref{keygen}a illustrates the key generation pipeline. The input challenge draws from a pool of $N^M$ combinations defined by $N$ voltage amplitudes $V_n$ ($n\in \left[1,N\right]$) and $M\in \left[1,\frac{P(P-1)}{2}\right]$ electrode pairs $x_j y_k$ where $x_j$ and $y_k$ ($j,k\in \left[1,P/2\right]$) are input and output electrodes, respectively. A single challenge comprises voltage pulses applied to $M$ electrode pairs. The voltage pulse $v_{jk}(V_n)$ applied to the electrodes $x_j y_k$ comprises a periodic triangular waveform with voltage amplitude $V_n$ chosen from the available pool of voltages, a fixed sweep rate $\Delta V$, and step duration $\Delta t$. The collected response is the $M$-dimensional output current $i_{jk}(V_n)$ measured in each electrode pair $x_j y_k$.\\
Figures~\ref{keygen}b-d present experimental results of the QNR response to multiple challenges created from $N=11$ voltage amplitudes $V_n \in \left(0,10\right]$~V with a step duration $\Delta t = 27$ ms and varying sweep rate $\Delta V \in \left[0.005,0.1\right]$~V/step, applied to $M=3$ electrode pairs. We chose these values from the available capabilities of the probe station used for the experiments (see Methods). In agreement with the simulation results presented in Fig.~\ref{theory}, the QNR creates multidimensional current orbits represented by hysteretic loops that scatter randomly in space and never intersect (Fig.~\ref{keygen}b). As in theoretical predictions (Fig.~\ref{theory}d), the QNR output response becomes narrower when the voltage-step variation is slower (Fig.~\ref{keygen}c) and manifests non-zero current reducing for slower voltage sweeps that quantify the acquired charge capacitance of the material. The average power consumption of each readout channel is $0.07$~nW per single voltage pulse applied.\\
Figure~\ref{keygen}d illustrates the QNR response to a periodic triangular excitation with amplitude $V_n = 10$~V and period time $T=1$~s. This period is sufficient to discharge any residual capacitance from the chip. The measurement shows that the response is repeatable, oscillating within 2.7~\% on the average value of the current-voltage output. We observe repeatable results with the same range of oscillations for different choices of $V_n$ and sweep rates $\Delta V$. Although recent literature reported multilevel current-voltage switching with various configurations of phase change materials~\cite{wu2021programmable}, to the best of our knowledge, no existing work reported the response of Fig.~\ref{keygen}b-d with repeatable continuous modulation of parameters.\\
We implemented the software decoder with two fully connected networks: a feedforward network transforming current orbits into a latent space of features, followed by a 6-layer deep network to convert features into OTKs (see Supplementary Note I). We test the accuracy of the decoder prediction with a total of 10M parameters trained on 1331 OTKs and obtain 100\% accuracy in mapping the response of the system to OTK.\\
Figure \ref{keygen}e compares the performance of different security primitive technologies using a standard manufacturing independent metric, the bit density per feature size area $bit/F^2$, which measures the amount of information stored in a given area of an integrated circuit~\cite{rajendran2021application}. The higher the bit density per feature size area, the larger the density of devices produced with a given manufacturing process. Current techniques based on encoding two states in a single primitive unit cell possess a constant bit density per feature size area with the highest performance of 1~$bit/F^2$ \cite{yang2020machine}.\\
In the case of QNR with challenges expressed as $N_b$ bits of binary sequences, in contrast, the bit density per feature size area is $N_b\cdot \frac{P-1}{2}$ $bit/F^2$ and increases linearly with the number of $P$ electrodes (see Methods for computation details). In the experimental configuration presented in this work with $P=3$ and $N_b=4$, the bit density is 3.5 $bit/F^2$, representing a three-fold increase over the current best~\cite{yang2020machine}. Using the current wire bonding technique with a unit cell area of $50~\times~50$ $\mu$m$^2$ \cite{jaafar2022establishment}, the QNR reaches a bit density of 660 $bit/F^2$ in a footprint of 1 cm$^2$ (see Methods).\\  
Based on experimental data, we compute that the QNR system has a reliability of 99.6\% while exhibiting a unique bit response of 50\%. The obtained values are in the range of the best-performing hardware security primitives currently available~\cite{naveenkumar2022design}.

\subsection{Key validation}
Figure \ref{AI_attack}a shows the implementation details of the VA. The input of the VA comprises a vector $\mathbf{X}$ of floating-point numbers obtained by converting the Challenge-OTK into a binary string (Supplementary Note II). The vector $\mathbf{X}$ inputs to a network that recurrently applies a nonlinear and non-invertible user-defined transformation $R(\mathbf{X})$ to previously generated states $\mathbf{X}_{N}=R(\mathbf{X}_{N-1})$. To comply with the requirement for a sensitive system discussed in Fig.\ref{concept_image1}, we here choose $R(\mathbf{X})$ as a mixing chaotic map \cite{ott_2002} generalized for multiple dimensions (Supplementary Note III). By iterating the network for $N$ times, the map generates a vector of states $\mathbf{R}(\mathbf{X})=[\mathbf{X}_1,\mathbf{X}_2, ...,\mathbf{X}_{N}]$. The vector $\mathbf{R}(\mathbf{X})$ feeds a linear readout layer, which provides the output $\mathbf{Y}=\mathbf{W} \cdot\mathbf{R}(\mathbf{X})$. The readout uses weights $\mathbf{W}$ trained to reproduce the input $\mathbf{X}$ in the output $\mathbf{Y}=\mathbf{X}=\mathbf{W} \cdot\mathbf{R}(\mathbf{X})$ when $\mathbf{X}$ represents the Challenge-OTK pair. The floating-point vector $\mathbf{Y}$ converts back to a binary string following the inverse transformation described in Supplementary Note II.\\
We validate the performance of the VA for the pool of 1331 Challenge-OTK pairs with a total length of 360 bits per pair and obtain 100\% user validation accuracy for the VA with 5M chaotic states and 32K trainable parameters. We train $\mathbf{W}$ with a least square formulation based on singular value decomposition that can achieve 100\% user validation accuracy by generating enough states in the VA possessing non-vanishing singular values (Supplementary Note IV).
\subsection{Security analysis and system resilience}
\begin{figure*}
\centering
\includegraphics[width=0.99\textwidth]{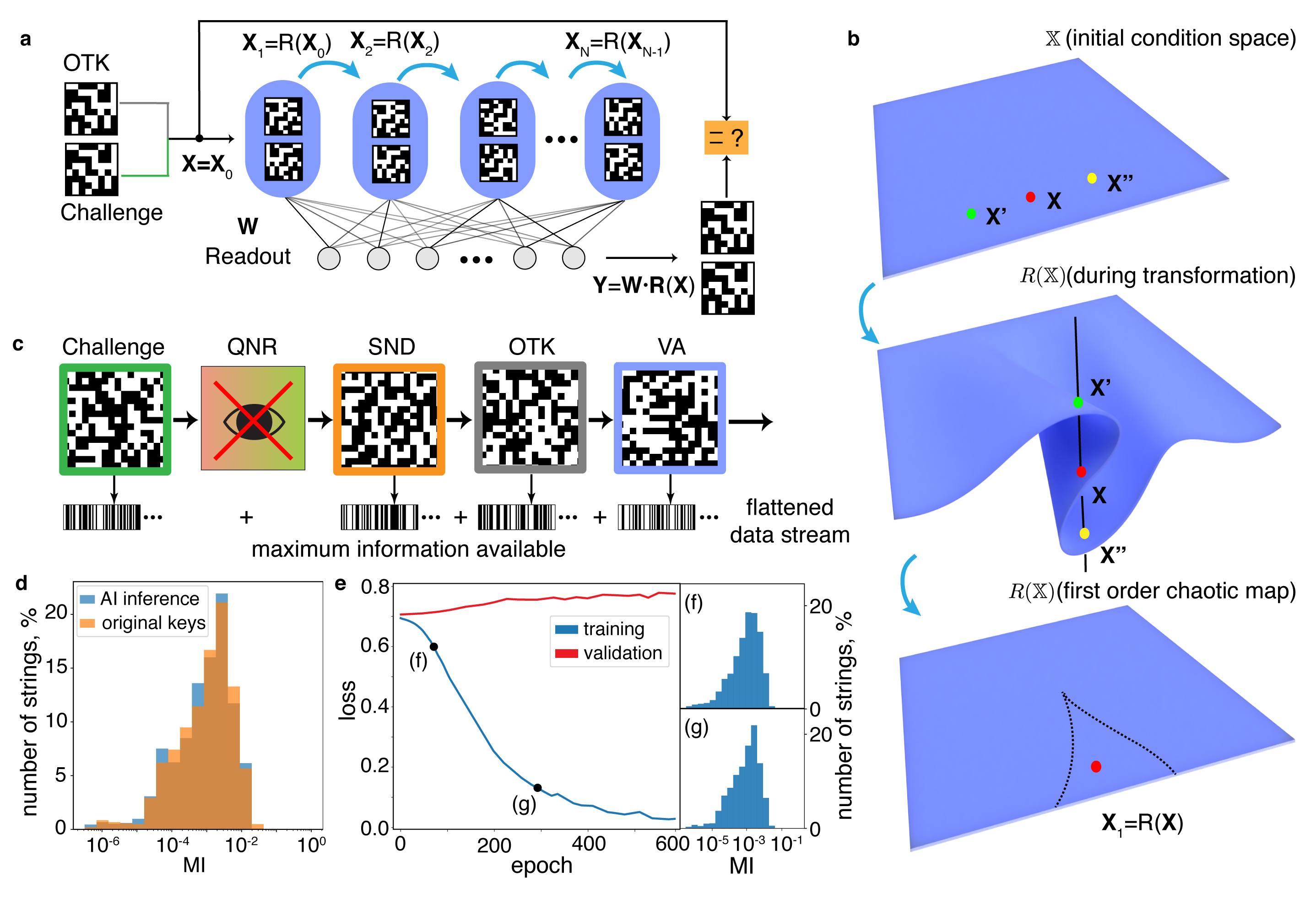}
\caption{\textbf{Key validation and resilience analysis} (a) VA implementation comprising a recurrent network applying a nonlinear and noninvertible transformation $R$ to the input OTK and a trained, fully connected readout layer. (b) Graphical illustration of the phase-space stretching and folding during the recurrent application of $R(\mathbf{X})$. (c) Inference attack diagram with an adversary who possesses a copy of all classical elements and collects the entire available flow of information. The information acquired, flattened as a single binary string, includes the input-output signals passing in each device and the status of all the classical components' internal variables. (d) Mutual Information (MI) between authentication OTK (orange area), and between correct and predicted OTK from an AI model trained on the complete sequence of data available from the transmission of $k-1$ OTK (blue area). (e) Training and validation loss of the AI inference network. (f,g) MI between AI-predicted and correct OTK after 40 and 300 epochs of training.}
\label{AI_attack}

\end{figure*}

Figure \ref{AI_attack}b summarizes the results on the robustness of the VA against an adversary who steals the VA and attempts to find an OTK. The adversary needs to solve equation $\textbf{X}=\mathbf{W}\cdot \mathbf{R}(\mathbf{X})$ for the vector $\mathbf{X}$ corresponding to a Challenge-OTK pair. As the matrix $\mathbf{W}$ is linear, the solution to this equation requires an inversion of the transformation $\mathbf{R}(\mathbf{X})$. However, the folding and stretching of the phase space of the chaotic map \cite{ott_2002} makes this operation impossible. Figure \ref{AI_attack}b illustrates this process for a set of 3 vectors with different coordinates $\mathbf{X}$, $\mathbf{X'}$, $\mathbf{X''}$ defined on a space of initial conditions. During the first iteration of the map, the deformation of phase space obscures the original location of these points (the deformed violet surface in the middle of Fig. \ref{AI_attack}b). When trajectories overlap due to folding (flat violet surface at the bottom of Fig. \ref{AI_attack}b), they lose the unique identification of their origins. As a result, a single overlapping output $\mathbf{X}_1=R(\mathbf{X})$ projects to multiple potential input values $\mathbf{X}$, $\mathbf{X'}$, $\mathbf{X''}$. Determining the specific initial condition that generated a given output is possible only by brute forcing a search for the possible initial conditions. Following the NSA recommendation for post-quantum resistant cryptography~\cite{NSA00}, using OTK lengths $>256$ bits makes the system computationally secure against brute force.\\
Figures \ref{AI_attack}c-g provide analysis results for the system resilience against statistical inference. We conduct this analysis using the ideas of Kerckchoff's principle, which establishes a strong form of security commonly used in cryptographic systems \cite{petitcolas2023kerckhoffs}. We consider an adversary who knows the system and who could copy every classical component except for the QNR. Cloning the QNR would require creating a material that shows the same feedforward network of Fig.~\ref{theory}e. This operation requires cloning the neural node transmission functions, which implies copying point-to-point the quantum dot nucleation dynamics that provide the same nonlinear response measured in Fig.~\ref{AFM}. Because HRTEM results in Fig.~\ref{TEM_image} demonstrate that quantum dots nucleate from amorphous regions, cloning the quantum scale phase transition process requires cloning the atomic structure of the material, ensuring that the electronic orbitals of the single atom that starts nucleation, or the atoms located at the edges of the quantum dot that grows in size, undergo the same transition as their neighbors. This process would violate the laws of quantum mechanics~\cite{Wootters1982-vn}, protecting this type of device against cloning.\\
In a single authentication instance, we assume an adversary intercepting all data available comprising the flattened binary OTK-challenge pairs, the output of the QNR, and the output and internal values of every classical element in the network, including the SND, and VA (Fig. \ref{AI_attack}c). We consider the scenario in which the adversary intercepts all but one $k-1$ authentication instance and attempts to infer the last $k$-th Challenge-OTK pair.
Figure \ref{AI_attack}d shows the distribution of MI between each authentication instance (orange bars). The average MI value is $10^{-3}$, which sets the maximal information gain that statistical inference can extract from the system. We illustrate this concept with an example using deep learning (Fig.~\ref{AI_attack}e-g) employing a 10M parameters network trained to learn the relationship between $k-1$ QNR responses and Challenge-OTK pairs from the flattened information flow. Every time, we consider a single key to predict from a network trained to learn the other keys' challenge-OTKs. We repeat this procedure on all OTKs available in the $1331$ set. The results of Fig.~\ref{AI_attack}e show that the network learns to reproduce the output of the QNR response after 600 epochs with training loss decreasing to values in the $10^{-2}$ range (solid blue line). However, the validation loss demonstrates that acquired knowledge is insignificant in predicting the new key, quantitatively observed as a constant, almost unitary loss (Fig.~\ref{AI_attack}e, solid red line). The results in Fig.~\ref{AI_attack}e-d show that the network learned the maximum amount of information available represented by the MI information among the keys (Fig.\ref{AI_attack}d,f-g solid blue area). However, because the MI is vanishing to $10^{-3}$, the inferred knowledge does not provide any gain to predict the following key. This example generalizes to any statistical inference, which does not provide any information gain because the MI available in this system tends to be zero.

\section*{Conclusion}
\noindent We implemented a 3-trillion nodes quantum-activated reservoir network integrated on-chip and demonstrated applications as large-scale AI hardware models for security primitives. The reservoir is quantum-protected against physical cloning and provides authentication keys with 99.6\% reliability and 50\% uniqueness. The information flows in the system have vanishing mutual information with average values of $10^{-3}$, making statistical inference attacks ineffective. The system does not require storing a challenge-key database using a validation autoencoder that reported 100\% user authentication accuracy.\\
The key generation capacity of the reservoir scales exponentially with the number of electrodes bonded to the chip. 
In the case of $P=24$ electrodes and the same challenge string encoding length of $N_b=4$ bits, a chip of 1 cm$^2$ can store $2^{1104}$ different keys. This capacity is $10^{100}$ times higher than the most extensive key space reported in an eFlash physical unclonable function~\cite{larimian2020lightweight}. Following the linear scaling law of CMOS circuits \cite{romli2015overview}, the energy efficiency of the chip with $50 \times 50$ $\mu$m$^2$ electrodes will reduce from current $4.2$ pJ/bit to a few fJ/bit. Manufacturing of the quantum neural reservoir is CMOS compatible, allowing chip integration into consumer electronics for the extensive panorama of applications beyond security, ranging from large-language models to large-scale transformer architecture for vision and video understanding. The quantum nucleation effects exploited in the reservoir follow an amorphous-to-amorphous transition in a nanosecond switching GST material~\cite{loke2012breaking}, enabling device operation bandwidth at Gb/s with available electronics.
\bibliography{bibliography}

\begin{methods}

\subsection{Fabrication of the QNR on-chip.}
We first clean a 4-inch silicon wafer in an ultrasonic bath and then dry it with nitrogen gas. We then oxidize the wafer in a Tystar tube furnace in a mixed atmosphere of oxygen and water vapor. Next, we use the wafer with silica film as a substrate for the deposition of GST film via the radio-frequency (RF) magnetron sputtering method. We then spin-coat AZ-ECI 3027 photoresist on it, followed by soft-baking at 100 °C. Next, we complete the exposure process in vacuum contact mode via a photolithography machine (EVG, EVG 6200) with an alignment mask. Subsequently, we immerse the exposed substrate in an AZ 726 MIF developer, developing the photoresist. After the development, we deposit metal electrodes on the patterned substrate using direct current (DC) magnetron sputtering. We then wire-bind the chip to a chip package (CSB02842) to obtain a final integrated QNR module.

\subsection{Sample characterization.}
We perform HRTEM on a Titan Themis-Z microscope by operating it at an accelerating voltage of 300 kV and with a beam current of 0.5 nA. We use a Protochips double tilt-heating holder to achieve an in situ annealing process during HRTEM measurements. We characterize amorphous GST films with a thickness of 20 nm using conductive atomic force microscopy (C-AFM, Dimension Icon AFM, Bruker). The probes (PFTUNA, Bruker) used in C-AFM are antimony-doped Si and coated with Pt-Ir. We apply two different measurement modes, DCUBE-TUNA and PeakForce-TUNA, to characterize the electrical properties of the GST thin film. We measure the current-voltage curves of varying electrode combinations of our QNR device at room temperature by using a manual probe station paired with a semiconductor characterization system (Keithley 4200-SCS).

\subsection{Bit density per feature size area.}
We assume the feature size $F^2$ of the QNR as the area of a single electrode bonded to the chip. By applying $N$ voltage amplitudes to $M=\frac{P(P-1)}{2}$ electrode pairs selected from the pool of possible combinations of P electrodes, we generate $N^M$ current orbits in the QNR. By using $N_b$ bits to represent each voltage amplitude, we obtain the total number of orbits supported by the QNR as $N_{orbits}=N^M=2^{\frac{N_b \cdot P(P-1)}{2}}$. This value corresponds to the length of $\log_2(N_{orbits})=N_b \cdot \frac{P(P-1)}{2} $ bits of the QNR binary output. We then obtain the bit density per feature size area $N_b\cdot \frac{P-1}{2}$ $bit/F^2$ by dividing the chip bit-length to the electrode area and the number of electrodes. For current wire bonding techniques with an electrode area of $50~\times~50$ $\mu$m$^2$ \cite{jaafar2022establishment}, the QNR chip contains approximately 330 electrodes for a similar electrode layout of input-output channels. With challenges encoded with $N_b=4$ bits, this configuration reaches a bit density of 660 $bit/F^2$ in the footprint of 1 cm$^2$.

\end{methods}

\begin{addendum}
 \item[Author contributions] A.~F. conceived and directed the work. Z.~H., N.~L. and F.~X. worked at GST material fabrication, characterization and chip integration. M.~S.~E. worked on the theory, including key generation, validation, and security analysis. A.~F., M.~S.~E. and Z.~H. wrote the manuscript. 
 \item[Competing Interests] The authors declare that they have no competing financial interests.
 \item[Correspondence] Correspondence and materials requests should be addressed to andrea.fratalocchi@kaust.edu.sa.
\item[Acknowledgments] The authors acknowledge Dr. Mohamed Ben Hassine and Dr. Long Chen for their assistance with TEM and AFM measurements.
 
\end{addendum}

\clearpage

\clearpage

\clearpage

\end{document}


\maketitle

\section*{Supplementary Note I - Software Neural Decoder}
\begin{figure*}[htb]
	\centering
	\includegraphics[width=1\linewidth]{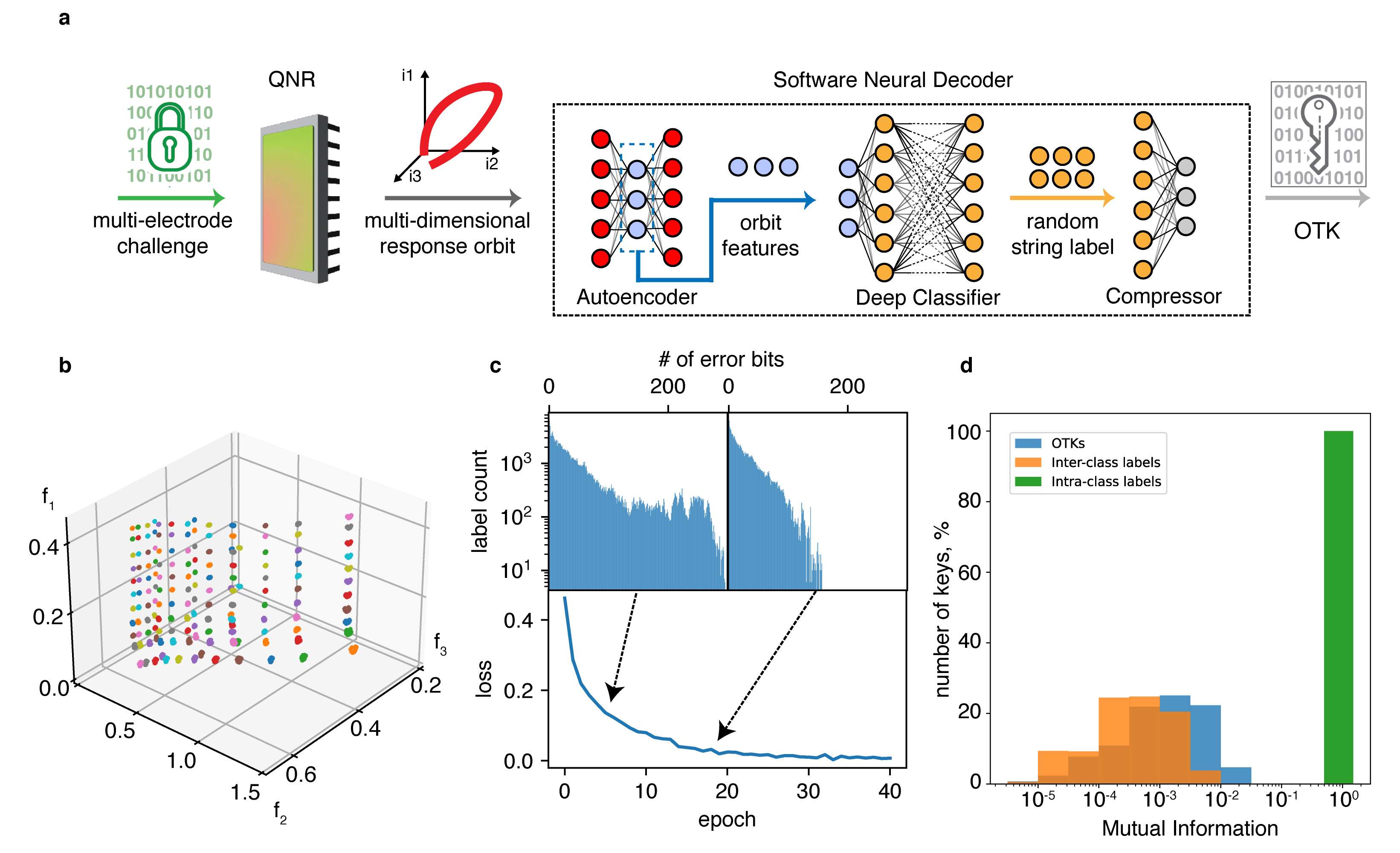}
	\caption{Training and performance of the Software Neural Decoder. \textbf{a} -- Key generation pipeline with expanded view on decoder components; \textbf{b} -- clustering of current orbit features in feature space; \textbf{c} -- Deep Classifier training; \textbf{d} - Intra-class and Inter-class Mutual Information between generated binary strings}
	\label{fig:S1}
\end{figure*}

Figure \ref{fig:S1}a shows the OTK authentication pipeline with an expanded view of the SND components. In the pipeline, we feed the multi-dimensional current response of the QNR $i_{jk}(V_n)$ to the autoencoder network. Its objective is to reduce the dimensionality of the data and extract unique features inherent to the distinct class of response orbits \cite{chen2014deep}. By extracting these features, we ensure that an identical challenge produces identical OTKs regardless of the oscillations of the electrical system. Next, we feed the orbit features to the trained deep classifier network. With this network, we match the features of orbit classes with the corresponding random binary string labels. Finally, a trained  network transforms labels into the OTK.\\
Figure \ref{fig:S1}b shows the results of orbit features extraction from experimental data. We use a dataset from 5 cycles of voltage sweeps performed on $M=3$ electrode pairs for network training. For each pair of electrodes, we apply the voltage pulse $v_{ij}(Vn)$ with amplitude $V_n$ in the range from 0.5 to 10 V with $N=11$ possible amplitudes. This results in 1331 classes of responses with 125 current orbits in each class. During training, we minimize the dimensionality of the autoencoder latent layer to identify non-crossing clustering in the feature space with a total of 1331 clusters. In Figure \ref{fig:S1}b, we plot 3-dimensional components $(f_1,f_2,f_3)$ of 8-dimensional features extracted from each current response to demonstrate the separability of classes.\\
Figure \ref{fig:S1}c illustrates the results of deep classifier network training. With an 8-dimensional input layer corresponding to the response feature space, it has 6 hidden layers with ReLU activation function and dimensionality determined by the number of response classes. We represent each class with random binary string labels. On the upper inset of figure \ref{fig:S1}c, we show the number of misclassified labels by the number of incorrect label bits after 5 and 20 training iterations, respectively. We observe that the number of error bits in the labels gradually decreases. With further training, the classifier learns to associate each response feature of each class with a correct label without any bit errors.\\
In Figure \ref{fig:S1}d, we show the distribution of Mutual Information (MI) between labels and OTKs. The MI between intra-class labels is unity as they are identical, while the average MI between the labels of different classes is $10^{-3}$. We generate OTKs from the compressor network trained to represent each label with a binary string in space with a dimensionality of the required key length (270 bits). With sufficient training, the label compression performs without bit errors. The distribution of MI between OTK pairs also suggests that there is no mutual information between these binary strings.

\section*{Supplementary Note II - Binary String Conversion Procedure}
Figure \ref{fig:S2}a illustrates the transformation process of a binary key string to the VA input layer. An original key combining challenge and OTK sequences has a total length of 360 bits (black-and-white pixel square), divided into eight channels, two for the challenge and six for the OTK. We encode each channel with a binary representation of a double-precision floating-point format, with the first bit of the string (sign bit) and the 44 first bits of the mantissa taking the values of a single channel string. We set 11 bits of the exponent at a zero offset, '01111111111', resulting in an encoded floating-point number that takes values in [-1;1]. We set the last 8 bits of the mantissa as '01111111', enabling binary string encoding with floating-point numbers without errors, thereby allowing for inference of the original key from the binary representation of these float numbers. With eight floating-point numbers for every channel (colored pixel rectangles), we represent an original binary string as an 8-dimensional initial condition at the input layer of a reservoir. We encode the entire training dataset of K keys as a matrix $\Omega$ with a size of $K\times8$.

\section*{Supplementary Note III - Chaotic Map of the VA}
\begin{figure*}[htb]
	\centering
	\includegraphics[width=0.85\linewidth]{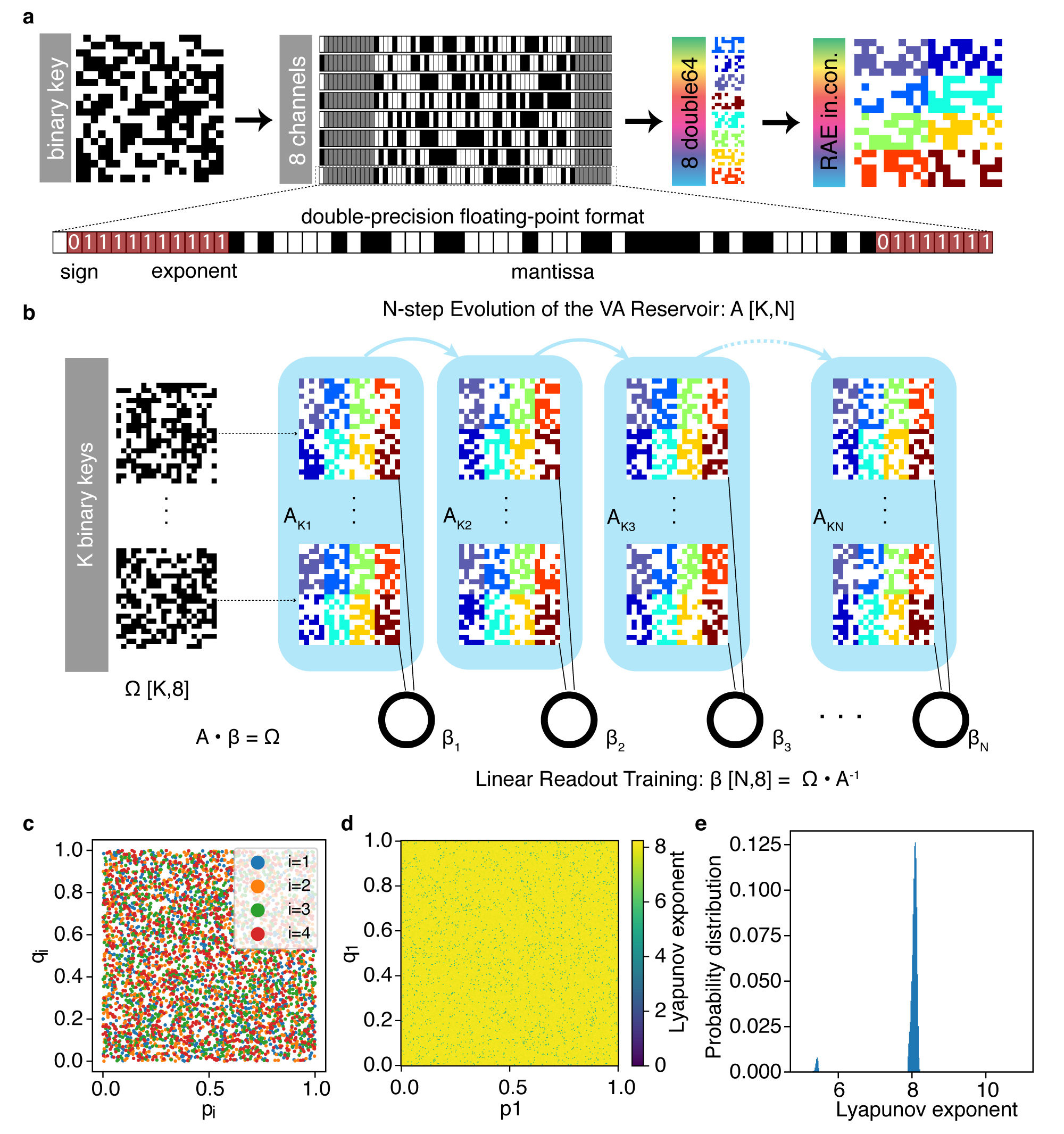}
	\caption{\textbf{a} -- Binary key to VA input layer transformation; \textbf{b} -- Validation Autoencoder network design; \textbf{c} -- evolution of chaotic reservoir states in coordinate space $(q_i,p_i)$; \textbf{d} -- Lyapunov exponent on the $(q_1,p_1)$ slice of the system; \textbf{e} -- distribution of Lyapunov exponent values on the $(q_1,p_1)$ slice.}
	\label{fig:S2}
\end{figure*}

Figure \ref{fig:S2}b visually represents the VA architecture and the training process. The VA consists of an input layer, a chaotic reservoir, and a linear readout. Each node in the input layer receives an 8-dimensional initial condition representing a binary string to be validated. The chaotic reservoir consists of the following Hamiltonian providing a multidimensional generalization to the two-dimensional kicked rotor model~\cite{ott_2002}:\\
\begin{equation}
    \label{hamiltonian}
    H=\sum_i^N \frac{p_i^2}{2}+\frac{L}{2\pi}\left [ \sum_i^N \cos(2\pi q_i)+\cos(2\pi((-1)^i q_i +q_{i+1})) \right ] \sum_i^N \delta (t-n\tau)
\end{equation}
with $p_i$ and $q_i$ are momenta and position coordinates, respectively, $L$ is a dimensionless parameter controlling the chaoticity of the system, and $N$ is the number of coordinate pairs with $q_{N+1}=q_1$. The Hamiltonian \eqref{hamiltonian} leads to the equations of motion:
\begin{equation}
    \label{ch_map}
    \begin{cases}
    q_i^{n+1}=(q_i^n+p_i^n)\mod 1
    \\
    p_i^{n+1}=p_i^n+L(\sin2\pi q_i^{n+1}+\sin2\pi\Delta q_{i+1}^{n+1}
-\sin2\pi\Delta q_{i}^{n+1})\mod 1
    \end{cases}
\end{equation}
with $p_i^n$ and $q_i^n$ represent the cordinate evolution at the $n$-th iteration, with $\Delta q_{i+1}^n=q_i^n-q_{i+1}^n$. When $L=0$, Eqs.~\eqref{ch_map} reduces to a set of uncoupled equations for $(q_i,p_i)$ pairs, with orbits fixed in the coordinate space. For increasing values of $L$, the system mixes different components and becomes progressively more chaotic. Figure \ref{fig:S2}c illustrates the evolution of a system with $L=100$ for 5000 iterations. All coordinate pairs have the same initial condition and exhibit coupled chaotic dynamics, with each $(q_i,p_i)$ pair having a unique trajectory. To measure the chaoticity of the system, we compute the Lyapunov exponent for the $(q_1,p_1)$ slice of the system. Figure~\ref{fig:S2}d shows that the exponent values are positive, indicating that the system is chaotic and highly sensitive to the initial conditions. With Figure \ref{fig:S2}e, we demonstrate the distribution of exponent values on this slice and point out that the chaotic reservoir possesses more than one largest Lyapunov exponent. This result indicates that the system is hyperchaotic with multiple degrees of chaotic behavior.
\section*{Supplementary Note IV - VA training}
To train the VA network, we initialize the reservoir with $K$ initial conditions corresponding to a training dataset's $K$ binary keys. Then, we discard the first ten iterations to ensure mixing and ergodic behavior. We evolve the system for a sufficiently large number of iterations $N$, with the evolution of the reservoir represented by the matrix $A$ of size $K\times N$. Each matrix row $a_i$ of size $1\times N$ contains $N$ chaotic reservoir states generated by an initial condition corresponding to a single binary string $k_i$. The linear readout beta is fully connected to the reservoir and has a size of $N\times 8$. Network training solves a linear problem of $A \times \beta = \omega$, where omega is the training set and beta is unknown. We apply the Singular Value Decomposition to the matrix $A = USV$, and with a large number of iterations $N$, the diagonalized matrix $S$ has eigenvalues of similar order. This approach minimizes the loss of numerical precision during the inversion of $S$. It maximizes the number of significant bits of a mantissa for key encoding. Finally, we compute the linear readout as $\beta = \omega  V^T  S^{-1}  U^T$ and fix it for the key validation process.

\printbibliography

